\documentclass[lnbip]{svmultln}
\usepackage{graphicx}
\usepackage{multirow}
\begin{document}
\frontmatter          
\pagestyle{headings}  

\mainmatter
%
%
\title{BPDMN: A Conservative Extension of BPMN with Enhanced Data Representation Capabilities}
\titlerunning{BPDMN: A Conservative Extension of the Business Process Modeling Notation}
\author{Matteo Magnani\inst{1} \and Danilo Montesi\inst{2}}
\authorrunning{Matteo Magnani \and Danilo Montesi}   
%
\tocauthor{Matteo Magnani (University of Bologna),
Danilo Montesi (University of Bologna)}
\institute{University of Bologna, Italy,\\
\email{matteo.magnani@cs.unibo.it}
\and
University of Bologna, Italy,\\
\email{danilo.montesi@unibo.it}}

\maketitle


\begin{abstract}
The design of business processes involves the usage of modeling languages, tools and methodologies.
In this paper we highlight and address a relevant limitation of the Business Process Modeling Notation (BPMN): its weak data representation capabilities. In particular, we extend it with data-specific constructs derived from existing data modeling notations and adapted to blend gracefully into BPMN diagrams. The extension has been developed
taking existing modeling languages and requirement analyses into account: we characterize our notation using the Workflow Data Patterns and provide mappings to the main XML-based business process languages.
\end{abstract}

\section{Introduction}

Today, the design of business processes often requires both managerial and technical expertise. For this reason, the Business Process Modeling Notation (BPMN) has been developed and then standardized (by the OMG) with the explicit aim of being understandable and usable by people with different roles and backgrounds, from top-level managers to IT personnel. The availability of a modeling language, together with tools and methodologies, is a key factor enabling the design of complex processes. However, currently this notation has a strong limitation, that we highlight and address in this paper: its weak data representation capabilities.

The BPMN specification states that data and information models \emph{are not part} of the notation, that data objects can be represented \emph{but} business process diagrams \emph{are not} data flow diagrams, and that data objects \emph{do not have any direct effect} on process flows \cite{BPMN}. Nonetheless, data objects have been defined as a predefined artifact of BPMN (together with comments and associations), because the BPMN specification itself declares that in some diagrams they represent \emph{the most important information to be modeled}.
For instance, one of the main advantages of Enterprise Resource Planning (ERP) systems is to provide unified access to the data. Master Data Management enables the interaction of different actors in a supply chain, to provide services like collaborative fulfillment networks.
Supply Chain Management systems deal with flows of materials, information and money, 
Product Lifecycle Management systems concern the transformation of materials into final products, and Customer Relationship Management systems manipulate customers' data to offer customized services. Basically, data is ubiquitous as far as business processes are concerned.

\begin{figure}
\centering
\includegraphics[width=.4\textwidth]{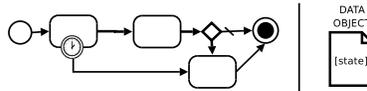}
\caption{On the left, an example of BPMN diagram with process-related constructs. On the right, the only visual construct used to represent data}
\label{fig:current}
\end{figure}

\begin{figure}
\centering
\includegraphics[width=.3\textwidth]{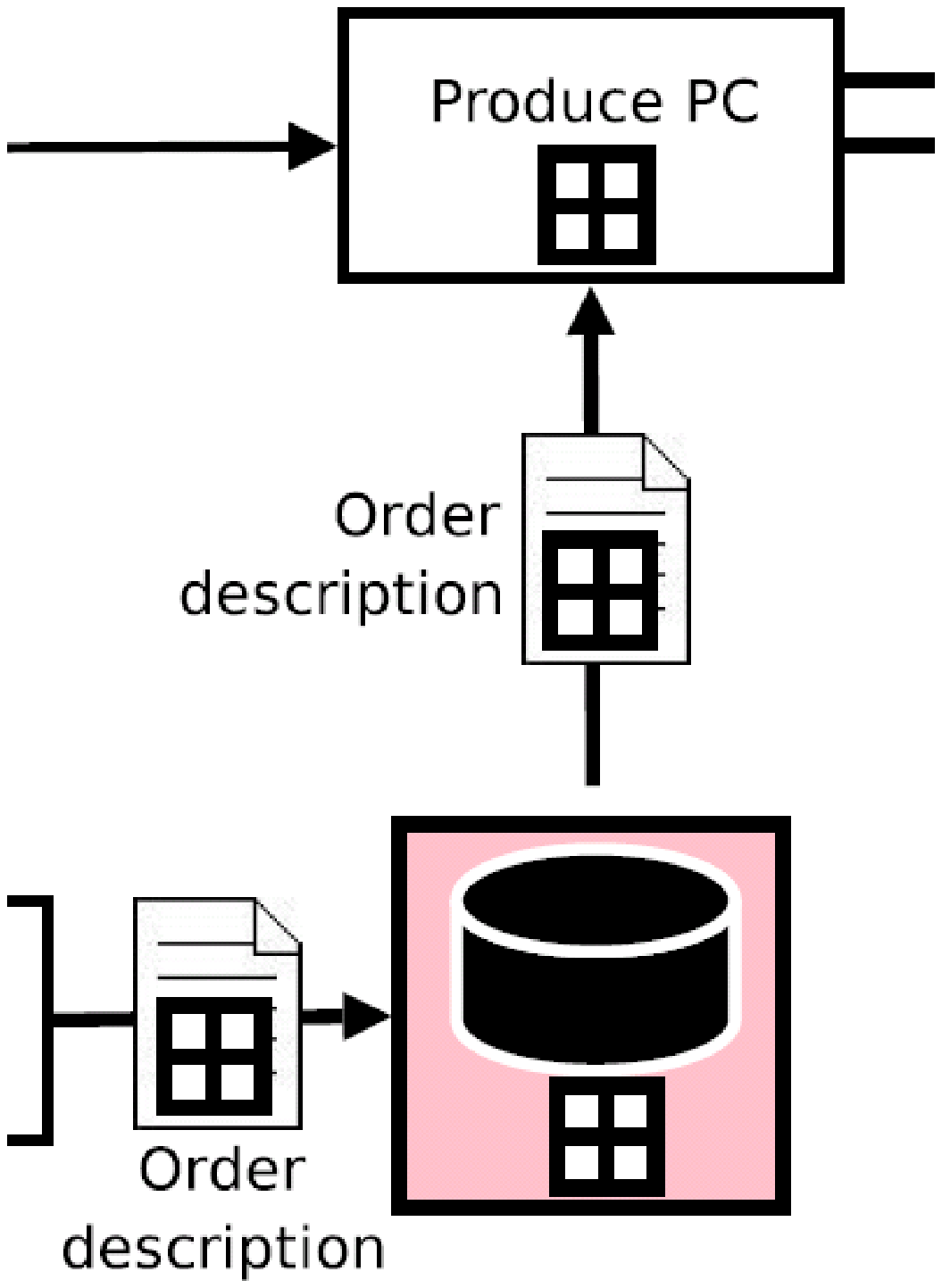}\hspace{1cm}\includegraphics[width=.3\textwidth]{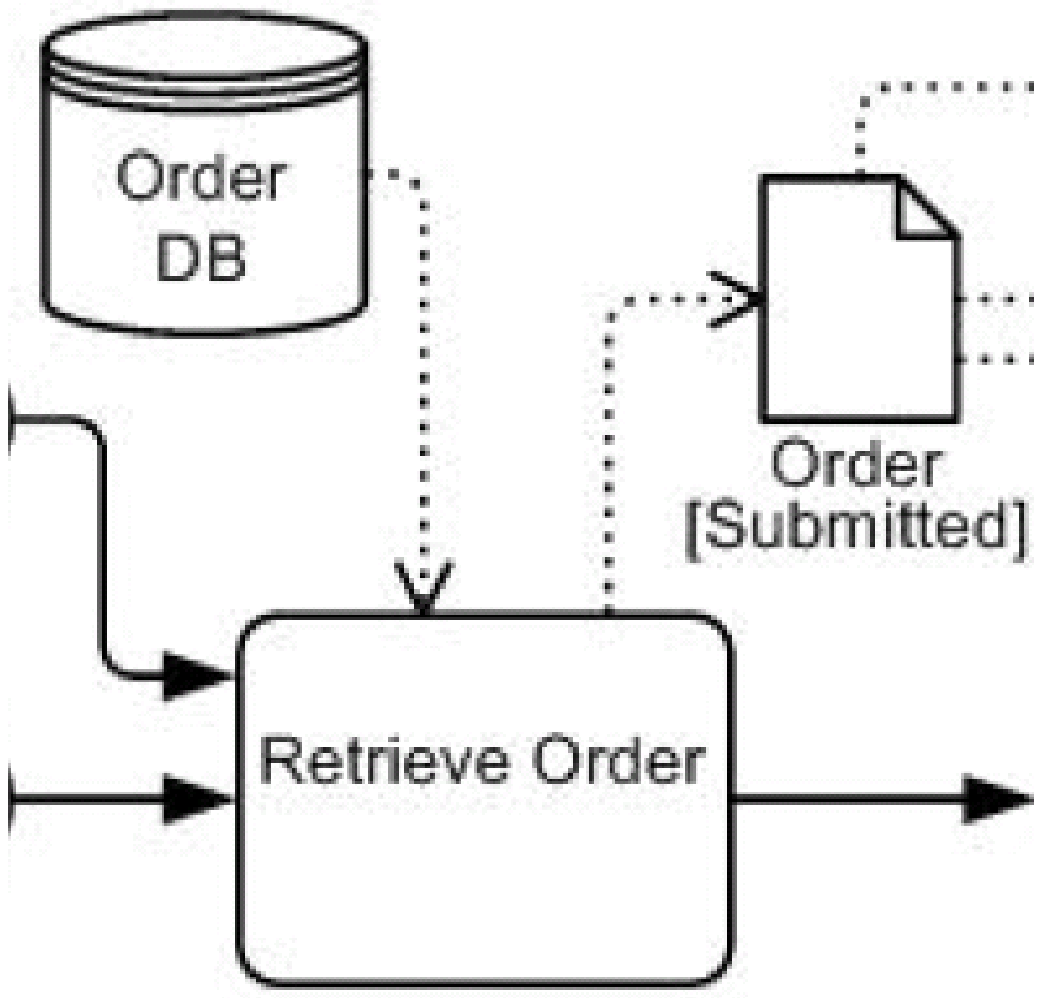}
\caption{Two snapshots from our 2007 report \cite{MagnaniTR07} (left) and from the current draft of the BPMN specification version 2.0 (right), showing a data store (DB) containing orders. In our extension it is also possible to expand data stores, as indicated by the + sign}
\label{fig:current2}
\end{figure}

\begin{figure}
\centering
\includegraphics[width=.3\textwidth]{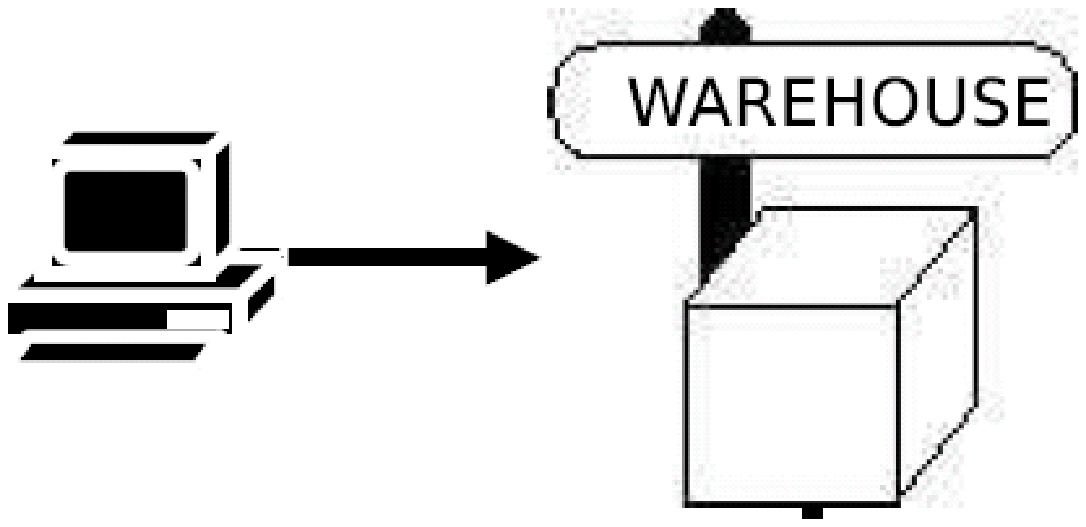}\hspace{1cm}\includegraphics[width=.3\textwidth]{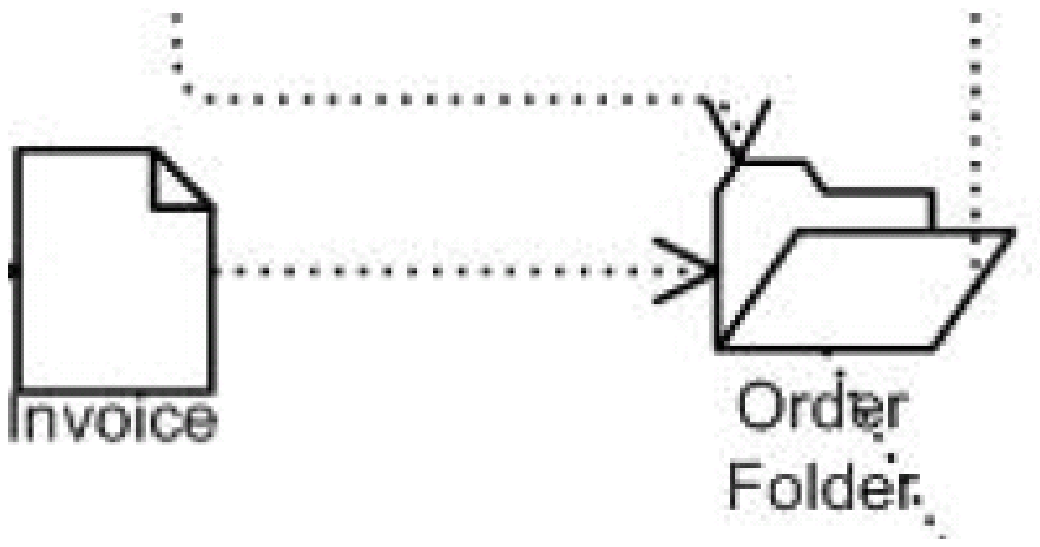}
\caption{Two snapshots from our 2007 report \cite{MagnaniTR07} (left) and from the current draft of the BPMN specification version 2.0 (right), showing custom icons to represent specific physical stores (a warehouse and a document folder, in these examples)}
\label{fig:current3}
\end{figure}

In BPMN, we can represent data using \emph{data objects} \cite{BPMN,White08}. Data objects are rectangles with a folded corner, where we can specify a state, like \texttt{approved} or \texttt{purchased}. If we compare this shape with the process-specific notation of BPMN (version 1.2), it clearly appears that information modeling capabilities are much less powerful and very informal, as illustrated in Figure~\ref{fig:current}. Some of the ideas presented in this paper and in its previous versions will probably be included in the forthcoming BPMN specification version 2.0, in particular the concept of data store (Figure~\ref{fig:current2}) and extended icons to represent alternative kinds of data objects/stores (Figure~\ref{fig:current3}). However, this document has not been released yet, and is not definitive, therefore in the following we will refer to the current official version of BPMN \cite{BPMN}.

In summary, data objects are \emph{too simple} to enable a proper representation of data, although data and processes are closely related in real business scenarios.
While it is reasonable that BPMN has been developed focusing on a single aspect of a business, i.e., processes, we think that data should not be modeled using a separate language, because we would not be able to represent the \emph{connection} between business processes and the data they transform. This is a very important aspect of an enterprise, because it is \emph{there} that new value is generated.

The main contribution of this paper is a conservative extension of BPMN, that we call Business Process and Data Modeling Notation (BPDMN). BPDMN has been defined as a result of three main activities:
\begin{itemize}
\item The \emph{mapping of real business processes} of a leading international mechanical company 
from informal descriptions to BPMN diagrams, as exemplified in Section~\ref{examples}. This has provided most of the requirements of the notation and motivated the development of a data-driven modeling methodology and a design tool supporting it, introduced in \cite{MagnaniICEIS09}.
\item The analysis of the data representation and manipulation capabilities of \emph{related business process management languages}, e.g., XPDL and BPEL, performed using existing requirement analyses. This motivated the inclusion of a \textbf{structure} into BPDMN data \textbf{objects}, and the definition of \textbf{data mappings}.
\item The analysis of existing conceptual \emph{data representation notations}.
\end{itemize}

\subsection{Structure of the paper, and what can be expected from it}

In the next section we introduce our notation, focusing on the new constructs (for details about other BPMN shapes the reader may consult \cite{White08}). In Section~\ref{mapping} we evaluate our proposal according to related works: we recall the notations that influenced BPDMN, we review our notation according to the Workflow Data Patterns, and indicate a mapping to the main XML-based languages used to represent and execute business processes. Then, in Section~\ref{examples} we present some examples of BPDMN diagrams. We conclude the paper with some final remarks. This work is part of a more general attempt to make BPMN a complete notation to support the design and the evaluation of business processes \cite{MagnaniBPM07,MagnaniICEIS09}.

It is worth noticing that, given the complexity of this topic, we do not claim or expect to be exhaustive. By the way, the BPMN specification itself misses many formal aspects, because we are dealing with a conceptual language whose flexibility and (sometimes) ambiguity may be used to adapt to different application contexts. However, we highlight the problem, introduce a candidate solution, provide an assessement of this solution and a comparison with other languages (using the well known data patterns), examples, and most importantly we expect to raise a discussion on a topic which is usually considered very important both by academics and pratictioners, according to our previous experiences. At the end of the next section, after having presented an overview of the notation, we discuss in more detail what has not been covered in the current version of our proposal, again with the aim of fostering discussions.

\section{BP\textit{D}MN}
\label{extension}

In this section we introduce the new notation. We first present its basic constructs, i.e., objects and stores, then we show how they are used inside diagrams (their dynamics). Finally, we introduce the idea of \emph{structured} stores and objects, allowing the representation of different levels of abstraction, and we provide some clarifications on what has not been included into the notation.

\subsection{Basic constructs (objects and stores)}

%

\begin{figure}
\centering
\includegraphics[width=.5\textwidth]{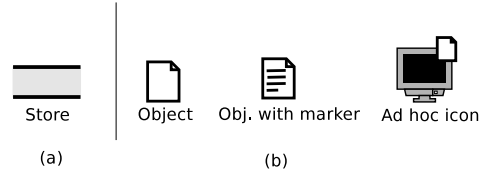}
\caption{Basic icons: (a) store and (b) extensions of the basic object icon}
\label{fig:extension1}
\end{figure}

\textbf{Stores} allow us to represent several facts about an enterprise, like that some information is available in the company databases, that some other information must be retrieved or bought by external information sources, or that products and documents (like invoices) are stored in specific locations and their insertion triggers other processes that can start working on them.
A store can represent a physical store, a digital store, or a set of variables that should be available throughout a process. 
The concept of \emph{store} and its icon have been taken from Data Flow Diagrams, as we have illustrated in Figure~\ref{fig:extension1}(a).


Stores contain \textbf{objects}, of which we can describe the dynamics --- how they are produced and manipulated.
Objects are used to represent generic physical and digital objects, like invoices, raw materials, products, or transactional data. The large number of entities that can be represented makes objects very powerful, but also potentially confusing and not self-understandable. Therefore, the first difference between objects in BPMN and in BPDMN is that we allow the usage of different icons to represent different kinds of objects (like UML stereotypes), with the aim of improving the readability of the diagrams \cite{UML}. This extensibility is very important, because icons that are intuitive in some contexts may not be the same in others, in addition vendors may wish to design specific icons for their products.
We define two ways to extend an object: either by adding a marker inside its basic icon, or by including a small basic icon on the top-right corner of the extension, as exemplified in Figure~\ref{fig:extension1}(b) --- this can also be used as a way to distinguish between digital and physical objects, without introducing yet another visual construct.

In the following section, we will introduce specific constructs to represent the life cycle (dynamics) of objects, i.e., \textbf{explicit  and implicit data flows}, and \textbf{data mappings}.


\subsection{Object dynamics}

\begin{figure}
\centering
\includegraphics[width=\textwidth]{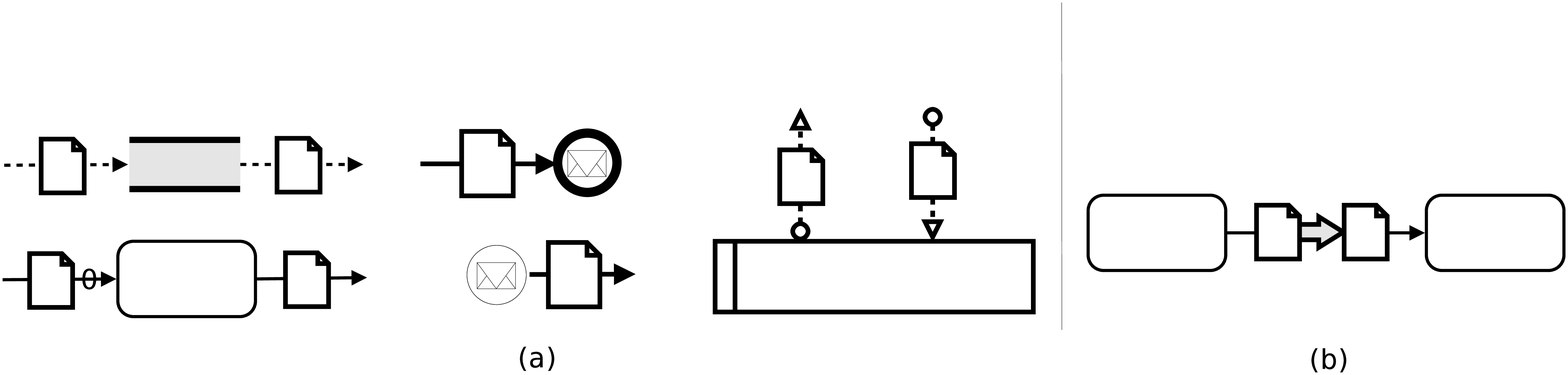}
\caption{Object dynamics: (a) object flows and (b) data mapping}
\label{fig:extension2}
\end{figure}

When objects are used inside BPDMN diagrams, where they do influence process flows, they must have a source and a target: they can be extracted from a store, produced by an activity, or received as a message from outside the diagram or from another pool, e.g., from a private process (Figure~\ref{fig:extension2}(a)). For example, a document can be taken from an archive, it can be produced by a \texttt{Write Document} task, or it can be received by mail from a customer (pool). Similarly, objects can be inserted into a store, sent to an external recipient or to another pool in the diagram, or provided to another task. In these cases, object icons are added on top of the corresponding connectors, as illustrated in the figure, and a small ``$0$'' on the corresponding flow indicates when the input is optional --- this is often the case when objects represent documentation or handbooks, that may or may not be necessary to the performers of the activity. Notice that objects can be exchanged between different pools through messages, supporting choreography.

\textbf{Data flows} naturally extend control flows (traditional BPMN flows): without data objects, the activity with the incoming flow must wait for the previous activity to finish, while with data objects it must wait for the previous activity to finish \emph{and} for the input data represented by the object (except when it is indicated as optional). We can also represent an \textbf{explicit data flow} using a dashed line, as in the top left corner of Figure~\ref{fig:extension2}. In addition, notice that objects are not restricted to single pools, but can be used to represent messages exchanged between different pools. With regard to BPDMN editors, we suggest to have an option to hide/unhide data objects on control flows, to provide the user a tool to focus on the data or process specific features of the diagram.

Finally, when an object is exchanged between two activities, it is possible that the output format of the first does not correspond to the input format of the other. This is frequent when processes are obtained as compositions of independent services, like in Service Oriented Architectures. In this case, we indicate the translation between the two objects using a solid arrow, as illustrated in Figure~\ref{fig:extension2}(b). This notation indicates a \textbf{data mapping}, and has associated metadata to store the mapping between input and output variables, like in BPEL --- we cannot provide additional details because of space limitations.

\subsection{Adding abstraction levels (structure) to stores and objects}

\begin{figure}
\centering
\includegraphics[width=.65\textwidth]{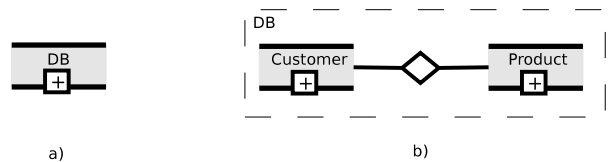}
\caption{Changing abstraction level with stores: (a) represents the same store of (b), at a higher abstraction level}
\label{fig:store_abs}
\end{figure}

Stores refer to physical or digital archives, or process variables when processes are executable. Stores can have a complex internal structure, and to represent it we will use the basic ER constructs: entities (i.e., stores), relationships (lines with diamonds) and generalizations (arrows). Each entity can be seen as a separate store, and different entities may be summarized using a higher-level entity at a different abstraction level, as we have exemplified in Figure~\ref{fig:store_abs}.
In this way, managers will have a few high level icons on their diagrams, 
while system administrators and programmers will unhide the entire data/object structures to implement them on the enterprise information system.

It is worth noticing that stores may become quite complex, as it happens in many large companies with very large databases and physical locations. However, this kind of information is usually not process-specific, but concerns many processes inside an organization. Therefore, we may expect that its models are already available somewhere, and must only be imported inside new diagrams.
For instance, schemata are already available in ERP systems, therefore we expect that many stores will not be designed inside process diagrams, but only linked to the process tasks that get or put objects from/into them. This concept can already be found in XPDL, where a \emph{package} is defined as a context which can contain several processes and make variables and resources available to all of them, without duplication and additional efforts.

In BPDMN, objects may refer to sets of variables, that will be associated to XML documents or WSDL messages at execution time. Therefore, also objects can be structured. In the same way, we can also associate a URL to an object, e.g., a link to an on line document.
Currently, in our tool we do not represent structures \emph{inside} the diagrams, but on a panel accessible by clicking on the object icon --- the internal structure of objects does not have to be connected to other graphical constructs. Objects can represent both global variables, when they are extracted from stores (in which case they will contain subsets of the store variables), or local variables, when they are generated by a task and sent to another through a sequence or message flow.

\subsection{Some final remarks about the notation}

Before introducing some example diagrams, it is important to clarify what BPDMN \emph{is not}.
In \cite{Sadiq04} the authors claim the importance of data modeling in business process languages to enable process validation. However, they also point out that it is not realistic to include ``all the relevant information on underlying databases'' into process diagrams. It is therefore necessary to find a compromise between the included features and the readability of the resulting diagrams. BPDMN enables the description of the dynamics (flow and manipulation) of data, but only up to a plausible level of detail, e.g., SQL queries and updates will not be visible on diagrams, but implicitly represented by BPMN activities like \texttt{Submit Form}. The typical approach used in statechart diagrams, i.e., representing how the state of an object changes, works only when objects are simple (like in BPMN). This is \emph{no longer possible} if we take into consideration complex data --- think of illustrating a database before and after an update!
In addition, although it is possible to simulate the behavior of BPMN diagrams to evaluate their cost and performance, consider that BPMN constructs describe a static view of processes, and not their real-time execution (e.g., there is not any concept of \emph{state} of the execution). Aspects like versioning of documents, which are very important, should not be dealt with in BPMN or BPDMN diagrams, but inside business process (workflow) execution systems.
Moreover, while data is a fundamental aspect of processes, there are specific features of the data that are independent of the peculiar process manipulating it, like consistency and replication. Similarly, access rights are an important feature of the data, but are usually not modeled at the conceptual level -- even if we do not exclude their inclusion into future extensions. We do not deal with methodologies to develop BPDMN (or BPMN) diagrams and cost analysis --- these are fundamental problems, out of the scope of our contribution and discussed separately in \cite{MagnaniBPM07,MagnaniICEIS09}. Finally, we do not provide a metamodel and a formal semantics for the notation: a standard meta model for BPMN is still under discussion, therefore it cannot be defined for our extended notation. Similarly, BPMN is not intended to be directly executable, and its behavior is currently described in the OMG specification using mappings to BPEL --- accordingly, we provide the mappings for our new constructs in Section~\ref{mapping}.

\section{Evaluation of BPDMN with regard to related languages}
\label{mapping}

\begin{figure}
\centering
\includegraphics[width=.5\textwidth]{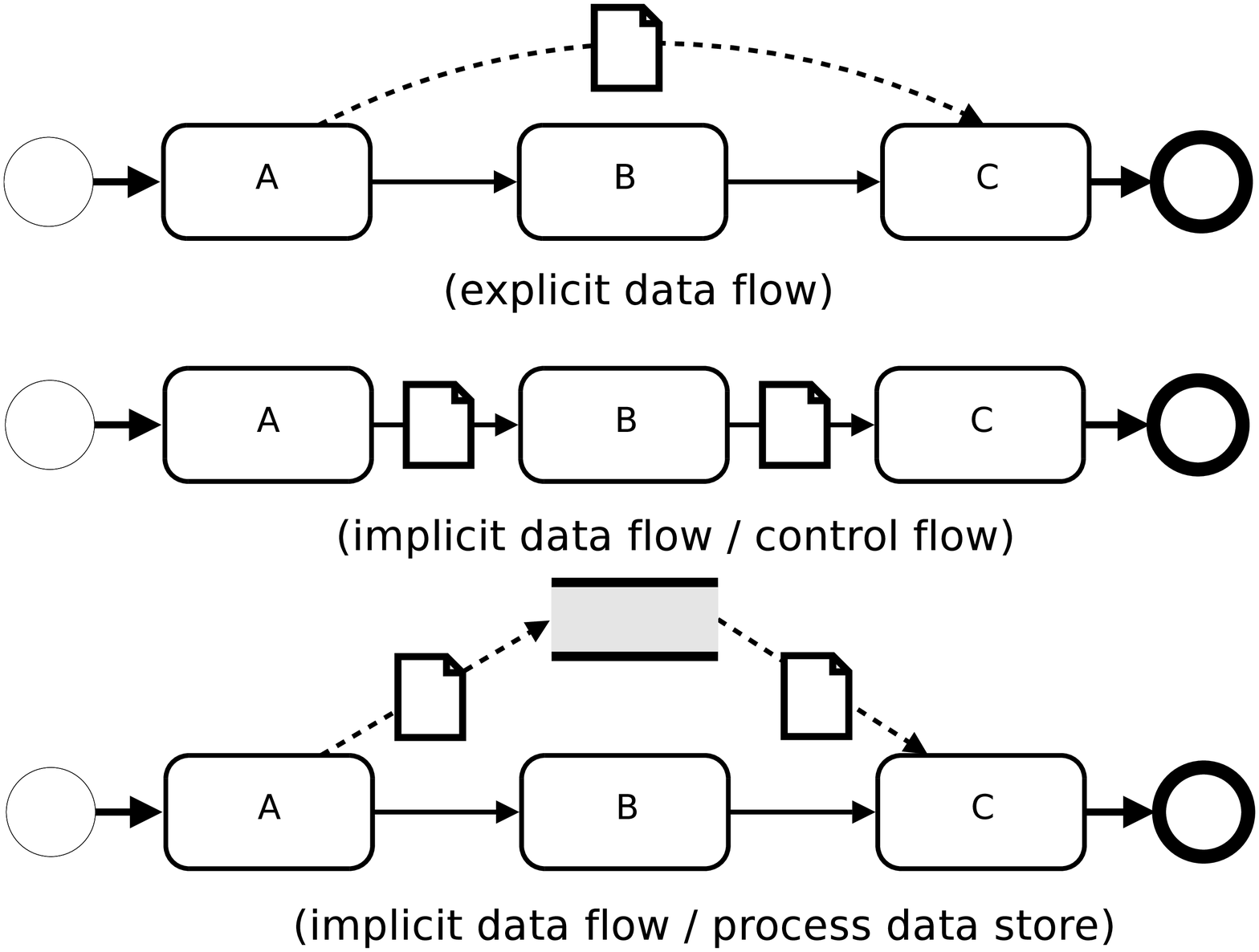}
\caption{Implementation models of data interaction in BPDMN (adapted from \cite{Sadiq04})}
\label{fig:implementation}
\end{figure}

BPDMN has been defined on the basis of existing requirement analyses. The requirements for a data and process modeling language identified in \cite{Sadiq04} regard the \emph{type}, the \emph{sources} and the \emph{structure} of data. In BPDMN, all the identified relevant data types (reference, operational, decision and contextual) can be represented using data objects, that belong to different classes depending on the context in which they are used, e.g., inside a data-based gateway (decision) or inside a sub-process (operational). We do not provide additional details here, because these aspects are already covered in BPMN. With regard to contextual data, the authors claim the necessity of representing complex structures, as in XML-based data exchange. Also this aspect is covered by our notation. Finally, all the three alternative \emph{implementation models} (as they are called in the paper) to exchange data inside a diagram can be modeled using BPDMN, as we have illustrated in Figure~\ref{fig:implementation}.

BPMN has not been defined as a competitor of existing workflow and business process management languages like UML 2.0 Activity Diagrams, YAWL, XPDL and BPEL \cite{Aalst05,UML,XPDL,BPEL}. On the contrary, it has been intended as an alternative and complementary language to be used by business analysts without technical knowledge, while UML was designed by and for software engineers, YAWL and BPEL provided respectively graphical and XML-based notations for executable processes and XPDL was intended as a portable exchange format. However, it is important to relate the data-specific capabilities of BPDMN to these languages. With regard to YAWL and UML Activity Diagrams, this can be done using the Workflow Data Patterns as a common criterion of comparison \cite{Russel04,Russell05} --- the analysis of these languages using the same patterns is reported in \cite{Russell06,Aalst05}. The relationship with the aforementioned XML based languages (BPEL and XPDL) will be discussed later in this section by indicating mappings from BPDMN to them, as it has already been done for BPEL in the BPMN specification \cite{BPMN}.

\begin{table}
\begin{center}
\begin{tabular}{|l|c|c||l|c|c|}
\hline
\textbf{data visibility} & 1 & 2 & \textbf{data interaction (cont.)} & 1 & 2 \\ 
\hline
1. Task data & + & + & 21. Env. to Case - Push & - & + \\ 
\hline
2. Block data & + & + & 22. Case to Env. - Pull & - & + \\ 
\hline
3. Scope data & - & - & 23. Workflow to Env. - Push & - & + \\ 
\hline
4. Multiple Instance data & +/- & + & 24. Env. to Workflow - Pull & - & + \\ 
\hline
5. Case data & + & + & 25. Env. to Workflow - Push & - & + \\ 
\hline
6. Folder data & - & - & 26. Workflow to Env. - Pull & - & + \\ 
\hline
7. Workflow data & - & + & \textbf{data transfer} &  & \\ 
\hline
8. Environment data & - & + & 27. by Value - Incoming & + & + \\ 
\hline
\textbf{data interaction (internal)} & & & 28. by Value - Outcoming & + & + \\ 
\hline
9. between tasks & + & + & 29. Copy in/Copy out & +/- & + \\ 
\hline
10. Block Tasl to Sub-wf Decomp. & + & + & 30. by Reference - Unlocked & - & + \\ 
\hline
11. Sub-wf Decomp. to Block Task & + & + & 31. by Reference - Locked & + & + \\ 
\hline
12. to Multiple Instance Task & - & + & 32. Data Transformation - input & +/- & + \\ 
\hline
13. from Multiple Instance Task & - & + & 33. Data Transformation - output & +/- & + \\ 
\hline
14. Case to Case & - & + & \textbf{data based routing} &  &  \\ 
\hline
\textbf{data interaction (external)} &  &  & 34. Task Precondition - Data exist. & + & + \\ 
\hline
15. Task to Env. - Push & + & + & 35. Task Precondition - Data val. & - & - \\ 
\hline
16. Env. to Task - Pull & + & + & 36. Task Postcondition - Data exist. & + & + \\ 
\hline
17. Env. to Task - Push & + & + & 37. Task Postcondition - Data val. & - & - \\ 
\hline
18. Task to Env. - Pull & + & + & 38. Event Based Task Trigger & + & + \\ 
\hline
19. Case to Env. - Push & - & + & 39. Data Based Task Trigger & + & + \\ 
\hline
20. Env. to Case - Pull & - & + & 40. Data-based Routing & + & + \\
\hline \hline
Structure & - & + & Explicit Data Flow & +/- & + \\
\hline
Data / Control Flow & +/- & + & Process Data Store & - & + \\
\hline
\end{tabular}
\end{center}
\caption{Comparison of data representation capabilities of (1) BPMN and (2) BPDMN, according to the Workflow Data Patterns and some additional requirements taken from \cite{Sadiq04}. + means supported, +/- partially supported, - not supported}
\label{tab:wdp}
\end{table}

We base our analysis on a previous examination of BPMN using the Workflow Data Patterns \cite{Wohed06}. In our opinion, these patterns are better suited to the analysis of languages at a lower abstraction level, like execution languages. In fact, they deal with aspects not covered by BPMN like data locking. For this reason, the analysis reported in \cite{Wohed06} is based on some assumptions about the underlying treatment of the data, detailed in the cited paper, that we adopt as well to enable a consistent comparison. The results of the analysis are reported in Table~\ref{tab:wdp}, and in the following we summarize the main findings --- the patterns indicated in the table are described in \cite{Russel04,Russell05}.
\begin{itemize}
\item With regard to the visibility of data inside blocks and cases (1,5), BPMN already supports it, but it is worth noticing that BPDMN supports it \emph{at the visual level}, and not only using attributes of the constructs. In fact, data visibility can be managed by putting a store at the appropriate level (sub-process, or diagram).
\item Patterns 3 and 6 are not supported, because they would need the extension of the \emph{group} construct of BPMN.
\item The majority of patterns concerning workflows and environment (7, 8, 19--26) are not supported in BPMN because it is not possible to represent them. On the contrary, stores may be used to represent external data, and therefore to model data accessible by the workflow engine or through the execution environment (like a shared database, or a set of system variables). Therefore, BPDMN adds support for these patterns.
\item BPDMN adds support for data management in multiple instance tasks (4, 12, 13). As suggested in \cite{Russel04}, this can be done by using a shared data storage to keep shared or instance-specific variables, which is allowed in our notation as exemplified in Figure~\ref{fig:implementation}.
\item Patterns 29--31 are made possible using stores to represent external or shared data repositories, but locking mechanisms cannot be explicitly enforced --- however, this is not a typical activity at the conceptual level.
\item Data Transformation (32, 33) is made possible in BPDMN through Data Mappings.
\item Finally, with regard to tasks 35 and 37 (imposing Pre- and Post- conditions to a task according to the value of a variable) it seems possible to enforce them using data-based gateways, but this would not be a specific feature of our extension. Therefore, as the authors of \cite{Wohed06} have regarded it as not possible in BPMN, we state the same about BPDMN.
\end{itemize}

XPDL and BPEL are relevant XML business process management languages with different main objectives (respectively, interoperability and execution) \cite{XPDL,BPEL}. In both languages there are elements corresponding to BPDMN constructs, that we have summarized in Table~\ref{tab:comp}.
%
%

\begin{table}
\begin{center}
\caption{A summary of mappings between BPDMN and BPEL/XPDL}
\label{tab:comp}
\begin{tabular}{lll}
\hline
BPDMN & BPEL & XPDL \\ 
\hline \hline
Store & \texttt{variable} & \texttt{DataField} \\ 
\hline
Object & Not available & \texttt{DataObject} \\ 
\hline
Object in/out from Task & \texttt{inputVariable} & \texttt{InputSet, OutputSet} \\ 
                        & \texttt{outputVariable} & \texttt{ActualParameters} \\ 
\hline
Object on Message & Not available & \texttt{Message} \\ 
\hline
Data Mapping & \texttt{assign} & \texttt{Assignments} \\
\hline
\end{tabular}
\end{center}
\end{table}

\section{Examples of BPDMN diagrams}
\label{examples}

In this section we present two examples of BPDMN diagrams. The first concerns an automated process defined using the BPEL language. The second is a real and typical example of a human-performed process, manipulating physical data (an instruction manual) but also interacting with a database. These two examples show how BPDMN can be used for both automated and non-automated processes, present cases where data modeling enhances the communication power of the diagrams, and illustrate how our extensions naturally blend into BPMN constructs, without any significant increase in the visual complexity of diagrams.


\begin{figure}
\centering
\includegraphics[width=\textwidth]{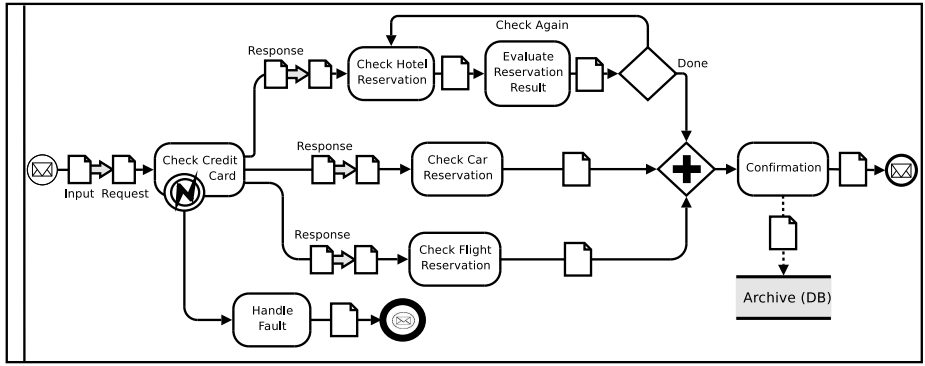}
\caption{A BPDMN diagram about travel booking (adapted and extended from \cite{White05})}
\label{fig:example2}
\end{figure}

In Figure~\ref{fig:example2} we have illustrated a BPDMN diagram corresponding to a BPEL travel booking process, designed using IBM WebSphere and previously used to exemplify a translation to BPMN \cite{White05}. At first glance, we can immediately appreciate the readability of BPDMN diagrams with respect to lower level languages, without losing the ability to represent data\footnote{To facilitate the work of the reviewers, we have reported the graphical representation of the original BPEL process in Appendix~\ref{app} --- this will be removed in case of acceptance of the paper}.
In this process, all data objects are digital documents (in particular, messages), and we have to apply the \emph{data mapping} feature of BPDMN to represent how values are passed through the diagram. In particular, expanding the data object originating from the \emph{message} start event we may see that it contains many variables (the \texttt{part} XML attribute in BPEL), e.g., \texttt{cardNumber}, \texttt{carCompany} and \texttt{hotelCompany}. If the credit card information is correct, then three services are invoked to check the hotel reservation, the car reservation and the flight reservation. Each service has its own input format, that can be seen expanding the data mapping before each of them. For example, the \texttt{Check Hotel Reservation} task requires a \texttt{name} attribute as input, and we must indicate that its value corresponds to the value of the \texttt{hotelCompany} attribute coming from the \texttt{Check Credit Card} task. This association is indicated in the data mapping. To further enrich the diagram, we have also added the registration of the travel plan into a database, indicated by the \texttt{Archive (DB)} store, to allow subsequent processes and users to retrieve information about current and past reservations.

\begin{figure}
\centering
\includegraphics[width=\textwidth]{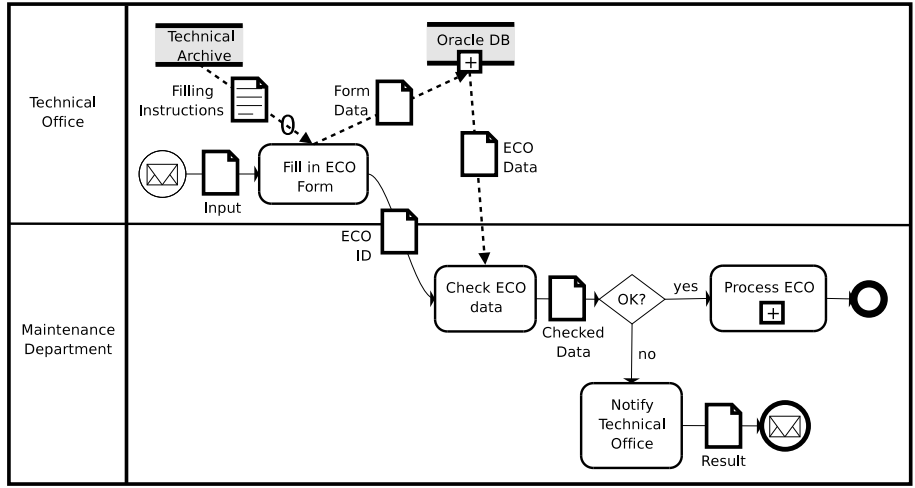}
\caption{A BPDMN diagram of an Engineering Change Order definition}
\label{fig:example1}
\end{figure}

Figure~\ref{fig:example1} represents part of a real business process regarding the upgrade of a mechanical device, and in particular the substitution of a component \cite{Clo}. This BPDMN diagram shows the definition of an \emph{Engineering Change Order} (ECO), i.e., the step in which the details of the upgrade are defined and checked. The process starts with the reception of a message with a request for an Engineering Change Order. From this request, the technical office prepares the data to be inserted into an Oracle database, filling in a form whose fields, like \texttt{component ID}, \texttt{device ID}, \texttt{replaced component}, and \texttt{procedure manager}, can be visualized by expanding the \texttt{Form Data} object. If needed, it is possible to consult a document with instructions on how to fill the form, that can be visualized by expanding the \texttt{Filling Instructions} object (this is in fact a link to a MS Word document available on line). Then, the ID of the order is sent to the maintenance department, that uses it to get the data from the database and verify it through the \texttt{Check ECO data} activity. If the maintenance department cannot process the ECO, e.g., if the new components are not available or do not conform with the specification of the device, then the technical office is notified and the process ends. Otherwise, the ECO can be processed.



\section{Conclusion}
\label{conclusion}
Documents, data, physical objects and several kinds of store are all fundamental aspects to determine the outcome of a business process.
In the BPEL specification, it is clearly stated that \emph{business processes include data--dependent behavior. For example, a supply--chain process depends on data such as the number of line items in an order, the total value of an order, or a deliver--by deadline} \cite{BPEL}. Therefore, in this paper we have identified some relevant constructs of the main existing formalisms to represent these aspects of a business, we have merged them into BPMN, and we have shown how these new visual constructs relate and can be mapped to other Business Process Management languages. We expect that the results of this effort and the refinements of our model that will be determined by practical applications and further dissemination will be an added value for business process designers, and a first step towards the definition of standard languages and tools to support design methodologies.

%
%

\appendix
\newpage

\section{Example diagram in BPEL (graphical notation)}
\label{app}

\begin{center}
\includegraphics[width=\textwidth]{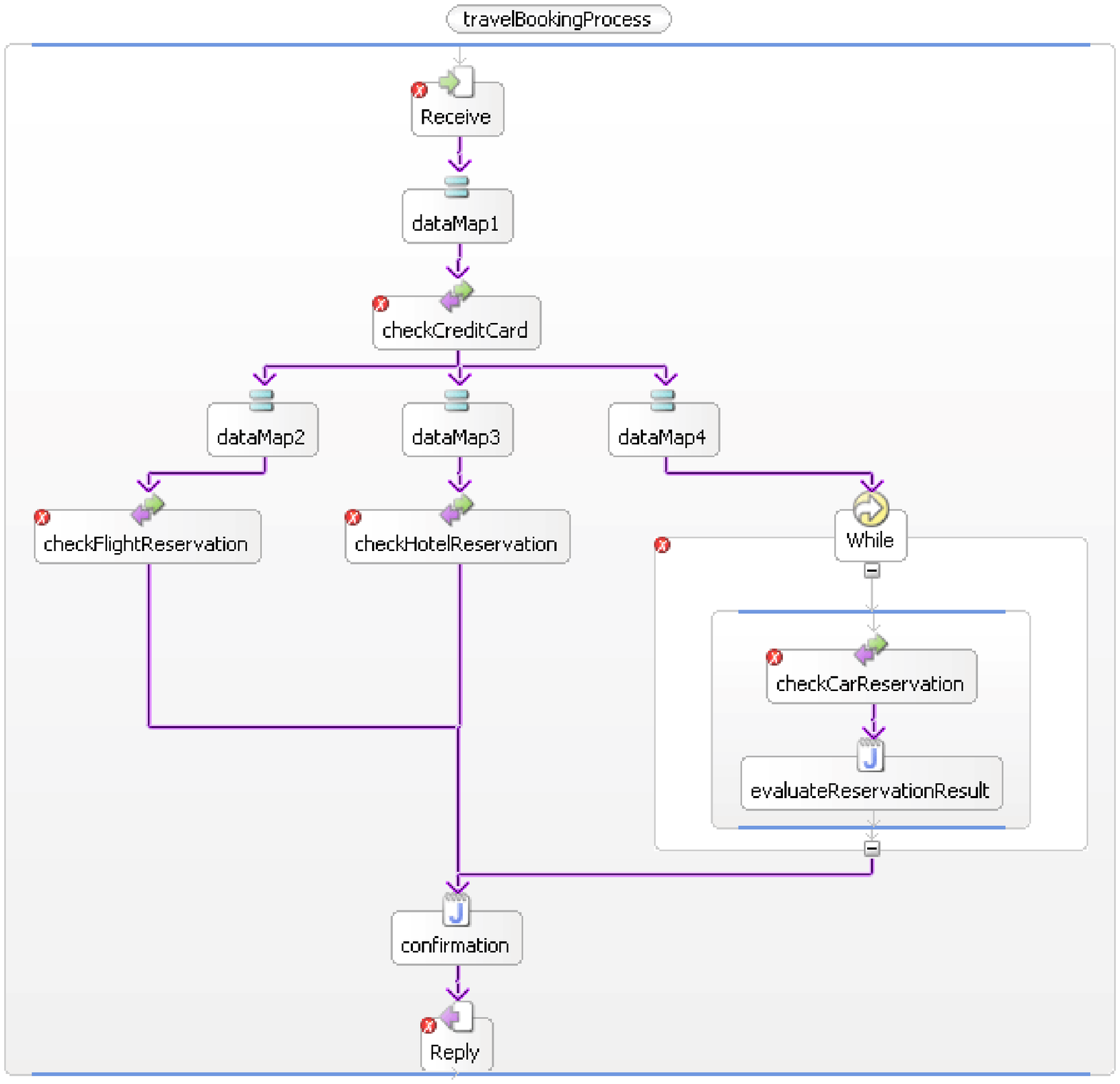}\\
A graphical view of a BPEL process about travel booking\\(source: WebSphere Information Center)
\end{center}

\section{Mapping between BPDMN, BPEL and XPDL}

In this section we show the relationships between BPEL, XPDL and BPDMN, as they have been summarized in Table~\ref{tab:comp}.

\subsection{Mapping to BPEL}

BPEL mainly concerns the automated execution of processes. With regard to BPDMN, stores and objects define which variables are used inside the process, objects indicate where these variables are used, and data mappings dictate how data is passed from one variable to another. In the following we will use the \texttt{Travel Booking} process as an example, and in particular we will show how the data received at the beginning of the process and stored into the \texttt{Input} object traverses the \texttt{Check Credit Card} activity --- then, the same translation can be applied to all subsequent activities.

Each store and object is translated into a set of \texttt{variable} elements. For instance, the \texttt{Input} and \texttt{Request} objects will be mapped to:
\begin{verbatim}
<variables>
   <variable name="input" messageType="input"/>
   <variable name="request" messageType="doCreditCardCheckingRequest"/>
   ...
</variables>
\end{verbatim}

In the previous BPEL fragment, \verb+input+ and \verb+doCreditCardCheckingRequest+ are references to WSDL message definitions, defining the parts and types of the variables --- we omit the details, that are not specific to BPDMN. At this point, a data mapping is used to map parts of the \texttt{input} variable to the input of the \texttt{Check Credit Card} activity, represented in the diagram by the \texttt{Request} object. A data mapping is directly translated into a BPEL \texttt{assign} element, where each expression corresponds to a \texttt{copy} element.
Although we allow complex mappings, looking at BPEL it appears that data transformations usually correspond to simple assignments (as witnessed by the names of the corresponding BPEL constructs).

\begin{verbatim}
<assign name="dm1">
   <copy>
      <from variable="input" part="cardNumber"/>
      <to variable="request" part="cardNumber"/>
   </copy>
   <copy>
      <from variable="input" part="cardType"/>
      <to variable="request" part="cardType"/>
   </copy> 
</assign>
\end{verbatim}

Now, the object can be used as input of the \texttt{Check Credit Card} activity (corresponding to a Web Service invocation), as follows:

\begin{verbatim}
<invoke name="Check Credit Card"
        inputVariable="request"
        outputVariable="response" ...>
        ...
</invoke>
\end{verbatim}
The code we have omitted inside the \verb+invoke+ element does not concern data, and can be seen in \cite{White05}.

\subsection{Mapping to XPDL}

In \textbf{XPDL}, with regard to stores, each variable can be mapped to a data field (following the usual rules of scope, so that if a store is defined inside a sub-process those variables will be defined in the corresponding sub-process section of the XPDL file). For example, in the \emph{ECO definition} process the Oracle database contains a \texttt{Device} table with a \texttt{deviceID} column, that will be mapped to the following code:

\begin{verbatim}
<DataField Id="OracleDB.Device.deviceID" Name="deviceID">
   <DataType><BasicType Type="STRING"/></DataType>
</DataField>
\end{verbatim}

Similarly, each object has a corresponding \texttt{DataObject} element, indicating the variables (data fields) to which it refers. The \texttt{ECO\_Data} object, which is extracted from the database and should thus refer to its data fields, will contain some of the variables defined in the \texttt{Oracle DB} store:

\begin{verbatim}
<DataObject id="ECO_Data" Name="Eco Data" ...>
   <DataFields>
       <DataField id="Device.deviceID" .../>
       <DataField id="Device.description" .../>
       ...
   </DataFields>
</DataObject>
\end{verbatim}

In addition, objects determine input and output of activities. In the translation to XPDL, differently from BPEL, we can represent both the graphical behavior, using \texttt{InputSet} and \texttt{OutputSet} elements with the identifiers of the objects, and the execution-related information, using \texttt{ActualParameter} elements to provide input and output to the corresponding application (which is defined elsewhere in the XPDL file, and not represented here). The following code corresponds to the \texttt{Check ECO Data} activity, and shows its input (\texttt{ECO\_Data}) and output (\texttt{Checked\_Data}) objects in addition to the data fields processed by the activity:

\begin{verbatim}
<Activity name="Check ECO Data">
   <InputSets><InputSet><Input ArtifactId="ECO_Data"/></></>
   <OutputSets><OutputSet><Output ArtifactId="Checked_Data"/></></>
   <Implementation>
      <Task><TaskApplication ...>
         <ActualParameters> ...
            <ActualParameter>ECO_Data.Device.deviceID</>
            ...
         <ActualParameters>
   </></></>
</Activity>
\end{verbatim}


Data mappings are translated into assignments among variables. In the working example, data mappings were not necessary, therefore to exemplify this case we may assume that the \texttt{Check ECO Data} activity has been automated, and that its associated Web Service requires a specific \texttt{Input} object. In this case, we should map the fields of the \texttt{ECO\_Data} object to the fields of the new \texttt{Input} object, as follows:
\begin{verbatim}
<Assignments> ...
   <Assignment>
      <Target>Input.device</Target>
      <Expression>ECO_Data.Device.deviceID</Expression>
   </Assignment>
   ...
</Assignments>
\end{verbatim}

Finally, objects can be associated to message flows, and mapped to the XPDL \verb+<Message>+ element --- in the same way as they are mapped when they are located on control flows.

%

\end{document}